\documentclass[conference]{IEEEtran}
\IEEEoverridecommandlockouts
\usepackage{cite}
\usepackage{amsmath,amssymb,amsfonts}
\usepackage{algorithmic}
\usepackage{graphicx}
\usepackage{textcomp}
\usepackage{xcolor}
\usepackage{comment}
\def\BibTeX{{\rm B\kern-.05em{\sc i\kern-.025em b}\kern-.08em
    T\kern-.1667em\lower.7ex\hbox{E}\kern-.125emX}}
\begin{document}

\title{Is my agent good enough? Evaluating Embodied Conversational Agents with Long and Short-term interactions
}

\author{\IEEEauthorblockN{Juliane B. S. dos Santos}
\IEEEauthorblockA{\textit{VHLAB} \\
\textit{School of Tech., PUCRS}\\
Porto Alegre, Brazil \\
juliane.beatrycce@edu.pucrs.br}
\and
\IEEEauthorblockN{Paulo Ricardo Knob}
\IEEEauthorblockA{\textit{VHLAB} \\
\textit{School of Tech., PUCRS}\\
Porto Alegre, Brazil \\
paulo.knob@edu.pucrs.br}
\and
\IEEEauthorblockN{Victor Putrich Scherer}
\IEEEauthorblockA{\textit{VHLAB} \\
\textit{School of Tech., PUCRS}\\
Porto Alegre, Brazil \\
Victor.Putrich@edu.pucrs.br}
\and
\IEEEauthorblockN{Soraia Raupp Musse}
\IEEEauthorblockA{\textit{VHLAB} \\
\textit{School of Tech., PUCRS}\\
Porto Alegre, Brazil \\
soraia.musse@pucrs.br}
}

\newcommand\red[1]{{\color{black}#1}}

\maketitle

\begin{abstract}
The use of digital resources has been increasing in every instance of today's society, being it in business or even ludic purposes.
Despite such ever increasing use of technologies as interfaces, in all fields, it seems that it lacks the importance of users perception 
in this context. \red{This work aims to present a case study about the evaluation of ECAs}. 
We propose a Long-Term Interaction (LTI) to evaluate our conversational agent  effectiveness through the user perception and compare it with Short-Term Interactions (STIs), performed by three users. Results show that many different aspects of users perception about the chosen ECA (i.e. Arthur) \red{could be evaluated in our case study}, in particular that LTI and STI are both important in order to have a better understanding of ECA impact in UX.
\end{abstract}


\section{Introduction}
\label{sec:introduction}

Embodied Conversational Agents (ECAs) can help people with some of their tasks while transmitting the sensation of talking with a real human. 
In addition, many games nowadays use intelligent virtual agents to improve the immersion and reality of the gaming experience, being it as Non-Playable Characters (NPCs) or tutorial/guides. It is important that people who play these games feel comfortable and enjoy such interactions; thus, it is equally important to measure and evaluate ECAs based on the perception of people.

When talking about UX, it is important to consider the user interaction with the proposed solution, software, or object as part of the development process or, even, while testing the final product. This interaction can happen in different ways, like using Short Interaction Sequences (SIS)~\cite{wimmer2010measuring} or by exposing users to the final product and asking for feedback at the end of the interaction~\cite{chen2010behavior,wimmer2010measuring}. 

In this work, \red{we present a case study which evaluates} an ECA based on users perception. First, we perform a Long-Term Interaction (LTI) and compare the perceptual results with Short-Term Interactions (STIs) performed by different users. The goal is to compare them and consolidate our recommendations for ECAs evaluation. 
The ECA used in our LTI, which occurred for 41 days, is Arthur, proposed by Knob et al.~\cite{knob2021arthur}. \red{In order to register the LTI, an user diary was built to log the daily interactions with Arthur, which we called Interaction Logbook.} 
In the end, with the results achieved, we provide improvement suggestions for the chosen ECA that can be useful for any development team of Embodied Conversational Agents.

\section{Related Work}
\label{sec:relatedWork}



Ruttkay et al.~\cite{ruttkay2004embodied} listed four of possible ECA behaviors: Embodiment, Input Recognition System, Model of personality and Emotions. In their model, the authors divide the ECA into three main modules: Embodiment, which is responsible for how the ECA will look like; Mental Capacities, which defines the personality and behavior of such ECA; and Application Interface, responsible for the information processing.
In the literature, it is possible to find many models proposed to build an ECA~\cite{zhang2019consistent,zhou2018emotional,knob2021arthur}. The work of Yalcin~\cite{yalccin2020empathy} aims to model empathetic behavior on Embodied Conversational Agents (ECAs). Their ECA has three stages: listening, where the agent captures input from the person it is talking to; thinking, where the agent process the information; and speaking, where the agent gives a proper response, both with words and gestural behavior.


According to Morville~\cite{10.1145/1187335.1187347}, seven factors influence User Experience (UX). Each of them is important to determine the success or failure of a project, being it a digital product or not. They are~\footnote{Please refer to Morville~\cite{10.1145/1187335.1187347} for further details}: Useful, Usable, Findable, Credible, Desirable, Accessible and Valuable.
Another concept that is vastly considered in UX is the “10 Usability Heuristics for User Interface Design”, created by Nielsen~\cite{nielsen1994enhancing}
They are~\footnote{For further information, please refer to Nielsen~\cite{nielsen1994enhancing} publication}: Visibility of system status; Match between system and the real world; User control and freedom; Consistency and standards; Error prevention; Recognition rather than recall; Flexibility and efficiency of use; Aesthetic and minimalist design; Help users recognize, diagnose, and recover from errors; and Help and documentation.

In addition to the Usability Heuristics~\cite{nielsen1994enhancing}, the Nielsen’s Severity Ratings~\cite{nielsen1995severity} are essential on the UX evaluation because it helps to understand the priority of each change that may be needed in the system. The ratings are: 0 = I don't agree that this is a usability problem at all; 1 = Cosmetic problem only: need not be fixed unless extra time is available on project; 2 = Minor usability problem: fixing this should be given low priority; 3 = Major usability problem: important to fix, so should be given high priority; and 4 = Usability catastrophe: imperative to fix this before product can be released.
Finally, there are some usability metrics used in the work of Ren et al.~\cite{castro2019usability} and Tullis and Albert~\cite{albert2013measuring}, that are based on the user perception such as Effectiveness, Efficiency and Satisfaction of the interaction with the product~\footnote{Please refer to Ren et al.~\cite{castro2019usability} and Tullis and Albert~\cite{albert2013measuring} for further details}.





Concerning ECAs evaluation, Bickmore et al.~\cite{bickmore2010maintaining} aimed to discover how to maintain user engagement in interactions with virtual agents. 
The results achieved demonstrate that user engagement can be manipulated using relatively simple techniques.
Babu et al.~\cite{babu2006would} evaluated Marve, a virtual receptionist placed at the entrance of their research laboratory. 
Results achieved indicate that the users enjoyed their interaction with Marve, also being able to perceive the agent as a human-like conversational partner.


\section{Arthur – the ECA}
\label{sec:methodArthur}


In order to conduct our research
we chose to work with Arthur, as proposed by Knob et al.~\cite{knob2021arthur}. 
The focus of Arthur is to provide a natural interaction with the final user by performing a consolidation of its Short-Term Memory and Long-Term Memory. Such process makes Arthur able to store and remember important pieces of the previous conversation
, creating the feeling that the user is having a natural interaction, as if it was with another human~\cite{knob2021arthur}. 
\subsubsection{Smalltalking Module}
\label{sec:methodST}

As defined by the Cambridge Dictionary, small talks have the basic definition that is a "conversation about things that are not important, often between people who do not know each other well"\footnote{https://dictionary.cambridge.org/dictionary/english/small-talk?q=small+talk}. The main advantage of bringing this concept to Arthur is that it allows to build a more approximate relation between virtual agent and human, especially when it is for a Long-term Interaction~\cite{10.1145/1963564.1963565}. 
In order to build our "Smalltalking" Module, 
we chose to create a simple conversational structure based on a Decision Tree~\cite{castle2020decision}. Thus, our "Smalltalking" Module is divided into three parts: Topics, Dialogues and Dialog Tree. 
For each of the Dialogues defined inside a certain Topic, a Dialog Tree is built. This Dialog Tree is composed of branches and nodes which define what Arthur can speak to the user. The path through the tree is made by Arthur saying/asking something to the user and waiting for his/her answer. Such answer is used to find the next utterance at the tree, by analyzing the polarity of the sentence (as done by Arthur~\cite{knob2021arthur}) and comparing it with the polarity of the next nodes. 
Finally, since the "Smalltalking" Module is triggered only when the interaction seems to "cool down", a timer was defined. We empirically defined that if the user says nothing to Arthur for 30 seconds, but keeps there, recognized by the Computer Vision module, Arthur randomly initiates a Smalltalk conversation.

\section{\red{The Case Study}}
\label{sec:method}

In this section, we present \red{the case study followed in this work}. 
Firstly, we conduct the Interaction Evaluation Steps, which are divided into three stages: \textit{1)} To conduct a Long-term Interaction with Arthur; \textit{2)} To conduct Short-term Interactions with a few users; and \textit{3)} To compare the findings and elaborate suggestions. The idea behind conducting the Short-Term Interactions was, indeed, to compare two different ways of interaction and see how useful each of them can be in our study. Next, we describe the proposal for evaluation of users perception, which is divided into two parts: \textit{1)} Heuristic Evaluation, following Nielsen's 10 Usability Heuristics~\cite{nielsen1994enhancing}; and \textit{2)} Severity Evaluation, following the Severity Ratings for Usability Problems~\cite{nielsen1995severity}. 


\subsection{The Long-term Interaction}
\label{sec:methodLTI}

Inspired by works like Kanda et al.~\cite{kanda2007two}, we evaluated Arthur using a Long-Term Interaction (LTI) with one user. \red{This person was contacted by our team and agreed to participate \red{in the case study presented here}. It was a woman, age of 24, with high experience with technology, but no experience with embodied conversational agents.} The main goal was to achieve measurable results on the perception of humans when interacting with Arthur. \red{Thus, the LTI user’s goal was to talk with Arthur everyday, as if Arthur was a friend of the user.} The LTI included the following tasks: \textit{i)} Perform daily interactions with the ECA model developed by Knob et al.~\cite{knob2021arthur}, as presented in Section~\ref{sec:methodArthur}; \textit{ii)} Register the interaction in the Logbook described in Section~\ref{sec:methodIS}; \textit{iii)} Evaluate the users perception based on the Effectiveness, Efficiency and Satisfaction metrics~\cite{albert2013measuring,castro2019usability}; and \textit{iv)} Evaluate the ECA following Nielsen’s Heuristics and Severity Ratings~\cite{nielsen1994enhancing, nielsen1995severity}.
Results are presented in Section~\ref{sec:results}. 

\subsubsection{Interaction Steps and Logbook}
\label{sec:methodIS}

Our Long-term Interaction 
had a total duration of 41 days. The approach chosen was to, each day, talk to Arthur as we talk to a person. The idea was to see if Arthur would be able to keep interacting in a natural way, as well as if he would be able to learn things \red{with/about the user}. 


To provide a detailed vision of how the Long-term Interaction with Arthur occurred, \red{a user diary was created, which we called Interaction Logbook.} 
It contained a few questions which were created based on the features of Arthur, but following the Effectiveness, Efficiency and Satisfaction criteria~\cite{castro2019usability} 
. In addition, a free text field was added to the questionnaire, so the user could write some details about her routine, feelings and so on. For each question, the researcher could select one of five options of a Likert scale: Very Unsatisfied, Unsatisfied, Regular, Satisfied, and Very Satisfied. After the LTI had finished, the same questionnaire was applied to the users selected to perform the STIs. This way, we were able to compare the results of both LTI and STIs in terms of extracted perceptual data from the users. These results are going to be presented in Section~\ref{sec:results}.




\subsection{The Short-term Interactions}
\label{sec:methodSTI}

We tested our ECA with three participants: A woman with medium familiarity with computers, 52 years old; a man with high familiarity with computers, 22 years old; and a man with medium familiarity with computers, 60 years old. \red{Neither of the three participants had any familiarity with conversational agents.} These were selected because they represent a great variability. \red{All three of them had a personal relation with the user who conducted the LTI, but not with the people who developed the project.} All the three participants interacted with Arthur and answered the same questions that the participant who conducted the LTI for 41 days. 
Each interaction occurred for about 10-15 minutes, where all users were asked to interact with Arthur as they pleased, reacting to what they found interesting, what bothered them, or what was challenging for them. \red{Nothing in special was asked: their only direction was to talk with Arthur as if they were talking with a new person they just met.}

\section{Results}
\label{sec:results}

In this section, we present the results obtained \red{in our case study}. 

\subsection{LTI 
assessment}
\label{sec:resultsLTIQuant}

The results were observed for all the 3 questions/topics:

    \textbf{Question 1 related to Effectiveness: It is defined as the accuracy and completeness with which users achieve specified goals in HCI. This should be responded considering Task completion, Accuracy and Recall~\cite{albert2013measuring,castro2019usability}}: For this question, the topics “Arthur’s effectiveness in learning something new” (Q1) and “Arthur’s effectiveness in showing something he learned” (Q2) were defined. For the most part of the LTI, these results were evaluated as Regular or Satisfactory, which means that Arthur was able to complete those actions.
    There were a few days where there was a bit harder to retrieve the information from Arthur or showing him something to learn, where the evaluation was Unsatisfied. 
    The obtained average of Likert scale for this question is 3.7 (considering Very Unsatisfied as 1 and Very Satisfied as 5), with a standard deviation of 0.61. 
    \textbf{Question 2 related to Efficiency: relates to the resources spent in relation to the accuracy and completeness with which the users achieve their goals. This should be responded considering Task completion time, Mental effort and Communication effort~\cite{albert2013measuring,castro2019usability}}: The topic “Effort applied while showing new things to Arthur” (Q1) was created to evaluate the way that the user can introduce new things to Arthur. In general, when the user says something like “Do you know my mother?”, Arthur tends to say “I do not know mother. Would you like to show me a picture?”. This topic was classified as Unsatisfied a few times, most of them for the first interaction days, as the user was not familiarized with the way Arthur works on this matter. After that, the interactions were evaluated as Regular, because Arthur begins to know many terms, making the interaction much easier. In this topic the average value obtained in the Likert scale was 3.01, with a standard deviation of 0.24. 
    \textbf{Question 3 related to Satisfaction: is defined as the degree to which user needs are satisfied when a product or system is used in a specified context of use. This should be responded considering ease-of-use, context-dependent questions, satisfaction before and during use, complexity control, physical discomfort of the interface, pleasure, the willing of use the chatbot again, and enjoyment and learnability~\cite{albert2013measuring,castro2019usability}}: Three topics were created to evaluate Arthur’s satisfaction: “Satisfaction with Arthur’s icebreakers” (Q1), “Satisfaction with Arthur’s answers to questions about him” (Q2) and “Satisfaction with Arthur’s answers to general questions” (Q3). In this case, the most chosen option was Unsatisfied due to some inconsistencies while interacting with Arthur, like the repetition of icebreakers and smalltalks. 
    The average value for Likert scale is 2.2, with a standard deviation of 0.38. 

\red{Moreover, we can take a look at how the ratings changed over time. Again, considering Very Unsatisfied as 1 and Very Satisfied as 5, Figures~\ref{fig:Effchanged} and~\ref{fig:Satchanged} present how the answers varied for Effectiveness Q2 and Satisfaction Q3, which were identified as the two extremes.

\begin{figure}[!htb]
  \centering
  \includegraphics[width=0.9\linewidth]{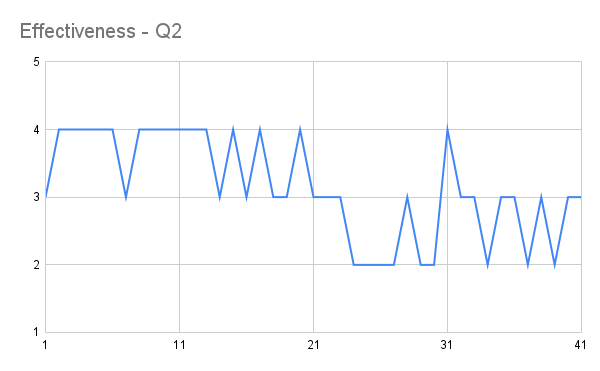}
  \caption{Ratings changing for Effectiveness Q2 (showing something that Arthur learned). A great variance can be seen in the answers as interactions went by.}
  \label{fig:Effchanged}
\end{figure}

Figure~\ref{fig:Effchanged} shows a great variance in the answers of the user concerning the second question of Effectiveness (Q2). It is possible to note that most answers were between Regular (3) and Satisfied (4) until about half of the interactions. Then, most answers were between Regular (3) and Unsatisfied (2), with one Satisfied (4). Based on this and the free text answers provided by the user, it seems that Arthur had little problem showing what he learned at the beginning. However, as he learned things and his memory became more complex, he began to find problems to show what he learned. We quote a part of one of the free text answers of the user: "...I showed him a picture of an anime, but when I try to retrieve it asking if he knows it, he shows a picture of himself, 
and if I only say the word anime, he only repeats it...". We believe that, given the many relationships that Arthur has built in its memory between all that he learned, he had some problem deciding which memory to retrieve when something was said to him, generating the discontentment.  

\begin{figure}[!htb]
  \centering
  \includegraphics[width=0.9\linewidth]{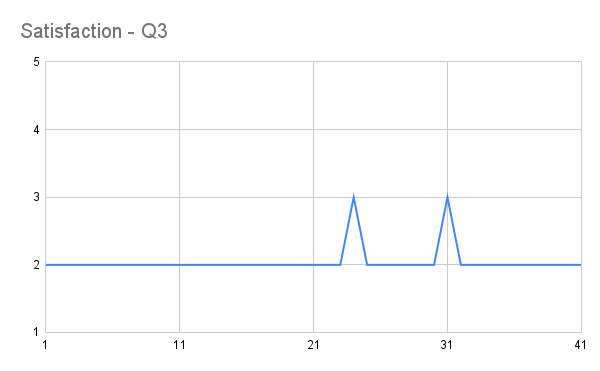}
  \caption{Ratings changing for Satisfaction Q3 (Arthur's answers to general questions). Almost no variance can be seen in the answers as interactions went by, with only two answers breaking the pattern.}
  \label{fig:Satchanged}
\end{figure}

On the other hand, Figure~\ref{fig:Satchanged} shows almost no variance in the answers of the user concerning the third question of Satisfaction (Q3). It is clear that most of the answers were Unsatisfied (2), with only two answers breaking the pattern as Regular (3). Once again, we quote a part of one of the free text answers of the user: "Arthur is kinda like saying random things. After saying Greetings, he started to say things like "Hum, show me some examples", or "I will check my agenda"...". It seems that Arthur found some problems when he needed to interact about something he had no knowledge about or was unsure on how to behave. Indeed, it could be also related with the situation reported in Figure~\ref{fig:Effchanged}: assuming that Arthur had some problem to decide which memory to retrieve, he would have problems to decide how to behave/answer as well.}

\subsection{STI 
assessment}
\label{sec:resultsSTIQuant}

\red{Although we are aware that a quantitative evaluation is not very useful having only three participants for the Short-Term Interactions, we believe it can be interesting to compare with the results achieved by the Long-Term Interaction.}
The three users were asked to interact with Arthur and, after that, answer the same questions used to compose the LTI Logbook. Some 
results are exposed as follows:

    \textbf{Question 1, regarding Effectiveness}: When looking at Arthur’s effectiveness in learning something new, all users were Satisfied or Very Satisfied, as they felt that Arthur was able to understand the things they said and showed to him. From the perspective of Arthur’s effectiveness, the obtained average of Likert scale for this question is 3.8 (considering Very Unsatisfied as 1 and Very Satisfied as 5), with a standard deviation of 0.47. 
    \textbf{Question 2, regarding Efficiency}: Although all users reported a difficulty on showing new things to Arthur (due to the time duration problem), all of them were Satisfied with this task. The obtained average of Likert scale for this question is 4, with a standard deviation of 0, 
    since all three participants answered as being Satisfied.
    \textbf{Question 3, regarding Satisfaction}: The satisfaction with Arthur had the most varied results in the STIs. The topic “Satisfaction with Arthur’s answers to general questions” had the best performance (all three subjects were Satisfied). On the other hand, the topic “Satisfaction with Arthur’s answers to questions about him” had answers varying from Unsatisfied to Very Satisfied. The obtained average of Likert scale for this question is 3.77, with a standard deviation of 0.57. 

Comparing results of LTI and STI, in Question 1, it is possible to see that while the user conducting the LTI evaluated Arthur's effectiveness as Regular most of the time, the users which conducted the STIs evaluated Arthur's effectiveness as Regular, Satisfied and Very Satisfied. 
Regarding the comparison of Question 2 results, 
provided by both LTI and STIs users, it is possible to notice that while the user conducting the LTI evaluated Arthur's efficiency as Regular most of the time, all the users which conducted the STIs were Satisfied with the efficiency. Finally, 
with respect to the Question 3, provided by both LTI and STIs users, it is possible to say that the user conducting the LTI evaluated its satisfaction with Arthur as Unsatisfied most of the time, the users which conducted the STIs evaluated their satisfaction with Arthur as Satisfied most of the time, even having a Very Satisfied answer.

\subsection{Usability Evaluation}
\label{sec:resultsInterfaceEval}

After having analysed LTI and STIs, we categorized every suggestion based on Nielsen’s 10 Usability Heuristics for User Interface Design~\cite{nielsen1994enhancing}, as well as Severity Ratings for Usability Problems~\cite{nielsen1995severity}, as it was found to be one of the best models for evaluating interfaces. The evaluation goes as follows:
    \textbf{1st Heuristic – Visibility of system status}: Arthur does not show the latest user message in the conversation field, which can give the impression that it was not correctly sent. 
    Some things can be done to solve this: the whole conversation can be shown in the text area; a notification sound can be triggered every time that the user sends a successful reply to Arthur; a visual notification that gives the impression that Arthur is thinking could be added.
    \textbf{Severity}: 3.
    \textbf{2nd and 3rd Heuristics – Match between system and the real world; User control and freedom}: These heuristics were not violated.
    \textbf{4th Heuristic – Consistency and standards}: The button used to send messages to Arthur is not intuitive and it is hard to understand (it is pictured as a button with only a "V" inside it). 
    The ideal option, in this case, would be to use a more standard button, like putting "Send/Send Message" on the button or use a button similar to the ones available in social media, which tends to be more comfortable and more used by people.
    \textbf{Severity}: 1.
    \textbf{5th and 6th Heuristics – Error prevention and Recognition rather than recall}: These heuristics were not violated.
    \textbf{7th Heuristic – Flexibility and efficiency of use}: Adding a faster and easier way for the user to send the messages to Arthur would be a good way to increase its efficiency. Instead of having to move a hand to the mouse and then click on the button, just pressing the “Enter” button on the keyboard would be better and more natural for the user.
    \textbf{Severity}: 2.
    \textbf{8th Heuristic – Aesthetic and minimalist design}: Arthur’s color palette is not a problem, but some components could be changed to make Arthur more aesthetic. For example: changing the format, position and size of the text box; buttons positioning; positioning of the webcam. Also, one of the comments specifies specifically the distribution of components on the interface.
    \textbf{Severity}: 0.
    \textbf{9th Heuristic – Help users recognize, diagnose, and recover from errors}: Even though rare, the errors displayed by Arthur could be better handled.
    \textbf{Severity}: 3.
    \textbf{10th Heuristic – Help and documentation}: Arthur does not have a “help” tool or documentation available for its interface. It would be interesting to add this kind of information in case of an eventual need to know more about Arthur’s functionalities in the future.
    \textbf{Severity}: 2.

In short, it is possible to notice that most of the suggestions were classified as low priority (0, 1 and 2 in Severity Ratings scale), with only two suggestions being \red{classified as a major problem} (3 in Severity Ratings scale) and none as a Usability Catastrophe (4 in Severity Ratings scale).

\section{Final Considerations}
\label{sec:conclusion}

This work discusses a case study performed to evaluate ECAs using Long-term (LTI) and Short-term Interactions (STIs). With the results achieved, we evaluated the users perception while interacting with Arthur, considering the effectiveness, efficiency, and satisfaction concepts for ECAs evaluation, described in the studies from Castro et al.~\cite{castro2019usability} and Tullis and Albert~\cite{albert2013measuring}. We included LTI and STIs in the methodology because, as we could see in our experiments, different periods of interaction can make users focus on different situations. By performing and comparing the LTI with the STIs, it was possible to notice that STI brought more satisfaction for the users at the end of the interactions and their qualitative comments focused more on concrete interface aspects. On the other hand, the LTI user "learned" how to use the ECA with the provided interface and focused more on communication aspects, such as ice breakers and small talks. So, we believe that both interactions are interesting and that is why we propose them both in our methodology to evaluate ECAs. 

As future work, we are planning to use the findings of this work to improve Arthur, focusing on the naturalness of interaction and on the UX evaluation. After that, we also plan to conduct other LTI and STIs in order to compare with the results of this work and evaluate if the changes indeed improved the behavior and the effectiveness of Arthur. \red{Moreover, it is possible that, with time, the participant who conducted the Long-Term Interaction (LTI) learned how to work around some aspects of Arthur which were causing issues. In this scenario, the later comfort with the ECA would be caused by such adaptation, not by improvements resulted from Arthur's learning or behavior. Our future work also focuses on these new interactions as a way to measure such possibility.

Concerning Short-Term Interactions, it is important to discuss some limitations which could have influenced the results. It is not possible to know what were the expectations of the three users about their interaction with Arthur, especially because none of them had any previous experience interacting with conversational agents. Therefore, it is possible that those three users perceived such interaction as an interesting experience and, even noticing some problems, it could also have been perceived as "part of the fun" of a new experience, which could have been reflected in the better evaluation perceived in the Results. In a future work, we plan to mitigate such possibility by conducting more STIs, including people which are already used with conversational agents.}

\bibliographystyle{IEEEtran}
\bibliography{bib}

\end{document}